\documentclass[a4paper]{jpconf}
\usepackage{graphicx}
\usepackage{amssymb}
\usepackage{amsmath}
\usepackage{acronym}        
\usepackage{xspace}         
\usepackage[hyphens]{url}   
\usepackage{cleveref}       
\usepackage{siunitx}        
\usepackage{booktabs}       
\usepackage{hepnames}       
\usepackage{fancyvrb}       
\usepackage{subfigure}

\DeclareSIUnit[number-unit-product = \;]{\permil}{\textperthousand}
\DeclareSIUnit[number-unit-product = \;]\byte{B}
\DeclareSIUnit[number-unit-product = \;]\flops{FLOPS}
\DeclareBinaryPrefix\kibi{Ki}{10}
\DeclareBinaryPrefix\mibi{Mi}{20}
\DeclareBinaryPrefix\gibi{Gi}{20}

\newcommand\picpath{plots}

\newcommand{\T}[1]{\mathrm{#1}}

\newcommand\Rcite[1]{Ref.~\cite{#1}}
\newcommand\ie{i.e.}
\newcommand\eg{e.g.}

\crefformat{section}{Section~#2#1#3}
\crefformat{figure}{Figure~#2#1#3}
\crefformat{table}{Table~#2#1#3}
\crefformat{chapter}{Chapter~#2#1#3}
\crefformat{appendix}{Appendix~#2#1#3}

\newcommand\lang[1]{\texttt{#1}}
\newcommand\prog[1]{\textsc{#1}}
\newcommand\code[1]{\texttt{#1}}
\newcommand\cd[1]{\code{#1}}
\newcommand\hardware[1]{{#1}}
\newcommand\order[1]{\ensuremath{\mathcal{O}\left(#1\right)}}

\newcommand\OMEGA{\prog{O'Mega}\xspace}
\newcommand\whz{\prog{Whizard}\xspace}

\newcommand\bytecode{byte-code\xspace}

\newacro{1POW}{one-particle off-shell wave function}
\newacro{APS}{antenna pole structure}
\newacro{BSM}{beyond the SM}
\newacro{CAS}{Computer Algebra System}
\newacro{CERN}{European Organization for Nuclear Research}
\newacro{CMF}{center of mass frame}
\newacro{CPU}{central processing unit}
\newacro{CSE}{commmon subexpression elimination}
\newacro{FPGA}{Field programmable gate array}
\newacro{GPU}{graphics processing unit}
\newacro{HL-LHC}{High Luminosity LHC}
\newacro{HMC}{Helicity Monte Carlo}
\newacro{ILC}{International Linear Collider}
\newacro{LCA}{leading color approximation}
\newacro{LHC}{Large Hadron Collider}
\newacro{LHS}{left-hand side}
\newacro{LO}{leading order}
\newacro{LSZ}{Lehmann-Symanzik-Zimmermann}
\newacro{MC}{Monte Carlo}
\newacro{MIC}{Many Integrated Cores}
\newacro{MSSM}{minimal supersymmetric SM}
\newacro{NLO}{next-to-leading order}
\newacro{NNLO}{next-to-next-to-leading order}
\newacro{NUMA}{Non-Uniform Memory Access}
\newacro{opcode}{operation code}
\newacro{OVM}{O'Mega virtual machine}
\newacro{PDG}{Particle Data Group}
\newacro{QCD}{quantum chromodynamics}
\newacro{QED}{quantum electrodynamics}
\newacro{RHS}{right-hand side}
\newacro{SIMD}{single instruction multiple data}
\newacro{SM}{standard model}
\newacro{VM}{virtual machine}

\usepackage[babel]{csquotes} 
\usepackage[style=numeric,
  maxnames=3,
  minnames=3,
  backend=bibtex,
  citestyle=numeric-comp,
  url=false,
  maxbibnames=5
  ]{biblatex}
\DeclareFieldFormat{journaltitle}{\emph{#1},}
\DeclareFieldFormat[inbook,thesis]{title}{{#1}\addperiod}
\DeclareFieldFormat[article]{title}{\mkbibemph{#1}}
\DeclareFieldFormat{eprint}{%
  \iffieldundef{eprintclass}
    {}
    {\thefield{eprintclass}/}%
    #1%
}
\DeclareFieldFormat{eprint:arxiv}{[\printfield{eprint}]}
\DeclareFieldFormat[article]{volume}{\textbf{#1}\addcolon\space}
\AtEveryBibitem{\clearfield{month}}
\AtEveryBibitem{\clearfield{title}}

\renewbibmacro{in:}{}

\begin{document}
\begin{flushright} DESY 16--034
\end{flushright}
\vspace{-5em}

\title{Making extreme computations possible with virtual machines}

\author{J Reuter$^1$, B Chokoufe Nejad$^1$, T Ohl$^2$}
\address{$^1$ DESY Theory Group, Notkestr. 85, D-22607 Hamburg\\
$^2$ University of W\"urzburg, Emil-Hilb-Weg 22, D-97074 W\"urzburg}

\ead{juergen.reuter@desy.de,bijan.chokoufe@desy.de,ohl@physik.uni-wuerzburg.de}

\begin{abstract}
State-of-the-art algorithms generate scattering amplitudes for high-energy
physics at leading order for high-multiplicity processes as compiled code (in
Fortran, C or C++).
For complicated processes the size of these libraries can become tremendous
(many GiB).
We show that amplitudes can be translated to byte-code instructions, which even
reduce the size by one order of magnitude.
The byte-code is interpreted by a Virtual Machine with runtimes comparable to
compiled code and a better scaling with additional legs.
We study the properties of this algorithm, as an extension of the Optimizing
Matrix Element Generator (O'Mega).
The bytecode matrix elements are available as alternative input for the event
generator \whz{}.
The bytecode interpreter can be implemented very compactly, which will help with
a future implementation on massively parallel GPUs.
\end{abstract}

\section{Introduction}
\label{s:intro}
Meta-programming is a popular approach in high-energy physics to determine the
expression of a cross section in a higher level programming language while the
numerical evaluation is performed in high-performance languages.
A problem, however, arises when the expression becomes so large that it is
impossible to compile and link, and hence to evaluate numerically, due to the
sheer size.
In \lang{Fortran}, which is known for its excellent numerical performance, we
typically encounter this problem for source code of gigabyte sizes irrespective
of the available memory.
In these proceedings, we sketch the ideas of \Rcite{1411.3834}, how to completely
circumvent the tedious compile step by using a \acf{VM}.
To avoid confusions, we note that a \ac{VM} is in our context a compiled
program, an \emph{interpreter}, that is able to read instructions, in the form
of \emph{\bytecode}, from disk and perform an arbitrary number of operations out
of a finite \emph{instruction set}.
A \ac{VM} allows the complexity of the computation to be only set by the
available hardware and not limited by software design or intermediate steps.
Furthermore, it is easy to implement and makes parallel evaluation obvious.

An important concern is of course whether the \ac{VM} can still compete with
compiled code in terms of speed.
The instructions have to be translated by the \ac{VM} to actual machine code,
which is a potential overhead.
However, in the computation of matrix elements a typical instruction
will correspond to a product of scalar, spinor or vector currents
which involves $O(10)$ arithmetical operations on complex numbers.
This suggests that the overhead might be small, which has however to be
proven by a concrete implementation.
In fact, we will show that a \ac{VM} can even be faster than compiled code for
certain processes and compilers since the formulation in terms of a \ac{VM} has
also benefits, especially for large multiplicities, as is discussed in detail
below.
More importantly, the runtime is in general in the same order of magnitude as
the compiled code and as such the \ac{VM} is very usable for general purpose
applications, where the clever use of \ac{MC} techniques can easily change the
number of points needed for convergence by orders of magnitude.
%
%
\section{General Virtual Machines}
\label{s:virtual_machine}
One might imagine a \ac{VM} as a machine, which has a number of registers, and
is given instructions, encoded in the \bytecode{}, how to act on them.
This picture is quite similar to a {CPU}, except that we are
doing this on a higher level, \ie{} our registers are arrays of \eg{} wave
functions or momenta and the instructions can encode scalar products or more
complicated expressions.
%
%
In order to use \acp{VM} for arbitrary processes, a dynamic construction is
desirable.
For such a construction, it is necessary to include a \emph{header} in the
\bytecode{}, which contains the number of objects that have to be allocated.
After this the body of instructions follows, whereby each line corresponds to a
certain operation that the \ac{VM} should perform on its registers.
The first number of an instruction is the \emph{\ac{opcode}} that specifies
which operation will be performed.
For illustration, consider the example \texttt{1~5~4~3},
which could be translated into
\code{momentum(5) = momentum(4) + momentum(3)}, a typical operation to compute
the $s$-channel momentum in a $2\,\to\,2$ scattering process.
Depending on the context, set by the \ac{opcode}, the following numbers have
different meanings but are typically addresses, \ie{} indices of objects, or
specify how exactly the function should act on the operands, by what numbers the
result should be multiplied, etc.
%
%
The interpreter is a very simple program that reads the \bytecode{} into memory
and then loops over the instruction block with a \code{decode} function, which
is basically a \code{select/case} statement depending on the \ac{opcode}.
The instructions can be instantly translated to physical machine code, compared
to the execution time of the relevant instructions, since the different types of
operations are already compiled and only the memory locations of the objects
have to be inserted.
The \bytecode{} file that is given to the interpreter completely dictates the
specific problem, or process in the cross section context, that should be
computed.
Input data or external parameters are given as arguments to the function call of
the \ac{VM}.
%
%
The generation of events for collider physics usually parallelizes trivially.
Since an integral is in most cases needed,
the same code is just evaluated multiple times with different input data.
The situation can change, however, for an extreme computation that already uses
all caches.
Depending on the size of the caches and the scheduler, evaluating such code
with multiple data at the same time, can run even slower than the
single-threaded execution.
Obviously, the computation is then so large, containing numerous objects, that
it is worth trying to parallelize the execution with a single set of input
data with shared memory.

The \bytecode{} lends itself nicely to parallelization as it can be split into
recursion \emph{levels}, whereby in each level all \emph{building blocks} are
non-nested and may be computed in parallel.
Different levels are separated by necessary synchronization points.
It is clear that one should aim to keep the number of synchronization points to
the inherent minimum of the computation for optimal performance.
This parallelization is straightforward in \lang{Fortran95/2003} and
\lang{OpenMP} and shown explicitly in \Rcite{1411.3834}.

As a side note, we mention that the sketched parallelization should be
very well suited for an implementation on a \ac{GPU}.
A common problem, encountered when trying to do scientific computing on a
\ac{GPU}, is the finite kernel size problem.
As noted \eg{} in \Rcite{MadGraphGPU2010}, large source code cannot be processed
by the \lang{CUDA} compiler, which is related to the fact that the numerous
cores on a \ac{GPU} are designed to execute simple operations very fast.
Dividing an amplitude into smaller pieces, which are computed one by one,
introduces more communication overhead
and is no ultimate solution since the compilation can still fail for complex
amplitudes~\cite{MadGraphGPU2010}.
The \ac{VM} on the other hand is a fixed small kernel, no matter how complex the
specific computation is.
A potential bottleneck might be the availability of the instruction block to all
threads, but this question has to be settled by an implementation and might have
a quite hardware-dependent answer.

Finally, we note that the phase-space parallelization mentioned in the beginning
of this subsection can still be applied.
When considering heterogeneous cluster or grid environments, where each node
is equipped with multi-core processors, a combination of distributed memory
parallelization for the combination of different phase-space points and shared
memory parallelization of a single point seems to be a quite natural and
extremely powerful combination.
\section{\OMEGA{} Virtual Machine}
\label{s:omega_virtual_machine}
The concept of a \ac{VM} can be easily applied to evaluate tree-level matrix
elements of arbitrary multiplicity.
The Optimizing Matrix Element Generator, \OMEGA{}~\cite{arXiv:hep-ph/0102195},
avoids the redundant representation of amplitudes in the form of
Feynman diagrams by using \acp{1POW} recursively.
\OMEGA{} tames herewith the computational growth with the number of external
particles from a factorial to an exponential one.

The model-independence is achieved in \OMEGA{} with the meta-programming ansatz
mentioned earlier whereby the symbolic representation is determined in
\lang{OCaml}.
This abstract expression is then translated to valid \lang{Fortran} code that is
automatically compiled and used in \whz{}~\cite{0708.4233,1206.3700,1112.1039}
for event generation.
We replace the last step now with an output module that produces \bytecode{}
instead of \lang{Fortran} code.

The number of distinct operations that have to be performed in the computation
of a cross section is related to the Feynman rules and therefore quite limited.
As such, these operations are very good candidates for the translation to
\bytecode{}.
In fact, this results in only about 80 different \acp{opcode} for the
complete \ac{SM}, which have been implemented in the \ac{OVM}.
For the parallel execution, we identify the different levels by the number of
external momenta a wave function is connected to or equivalently the number of
summands in the momentum of the wave function.
\section{Speed Benchmarks}
\label{s:benchmarks}
%
All processes shown here, and various others, have been validated against the
compiled versions where possible.
We stress that every process is computed in its respective model (QCD or SM) to
\emph{full} tree-level order including all interferences and we have not
restricted \eg{} the Drell-Yan amplitudes to only one electroweak propagator.
We use the term SM for the full electroweak theory together with QCD and a
nontrivial Yukawa matrix but without higher dimensional couplings like
$\PHiggs{}\to\Pgluon{}\Pgluon{}$.

\begin{figure}[htb]
\subfigure[Pure gluon amplitudes $2\to(n-2)g$]{\includegraphics[width=0.49\textwidth]{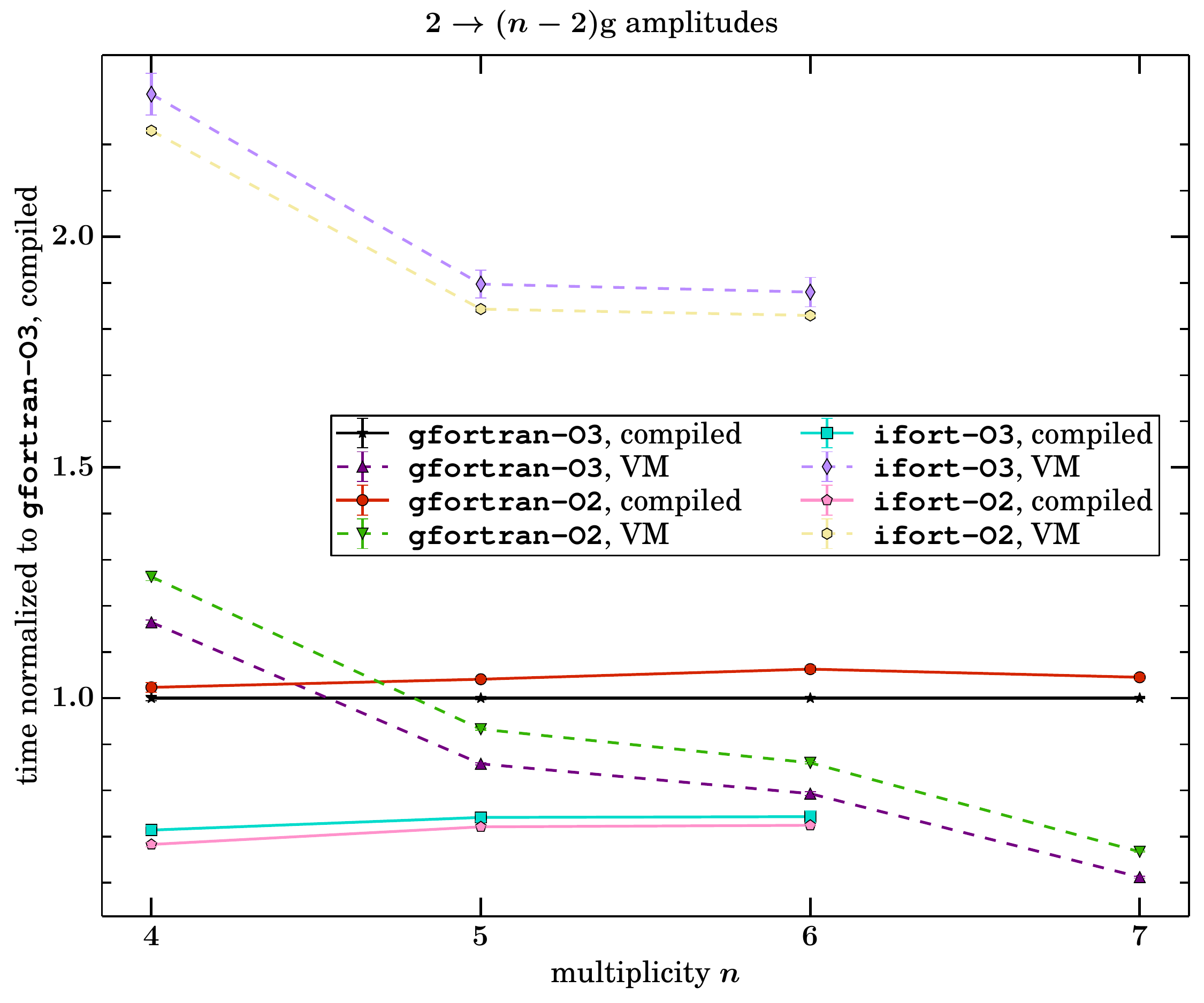}}\hfill
\subfigure[\ac{SM} Drell-Yan process $\Pqu \Paqu \to \Ppositron \Pelectron n j$]{\includegraphics[width=0.49\textwidth]{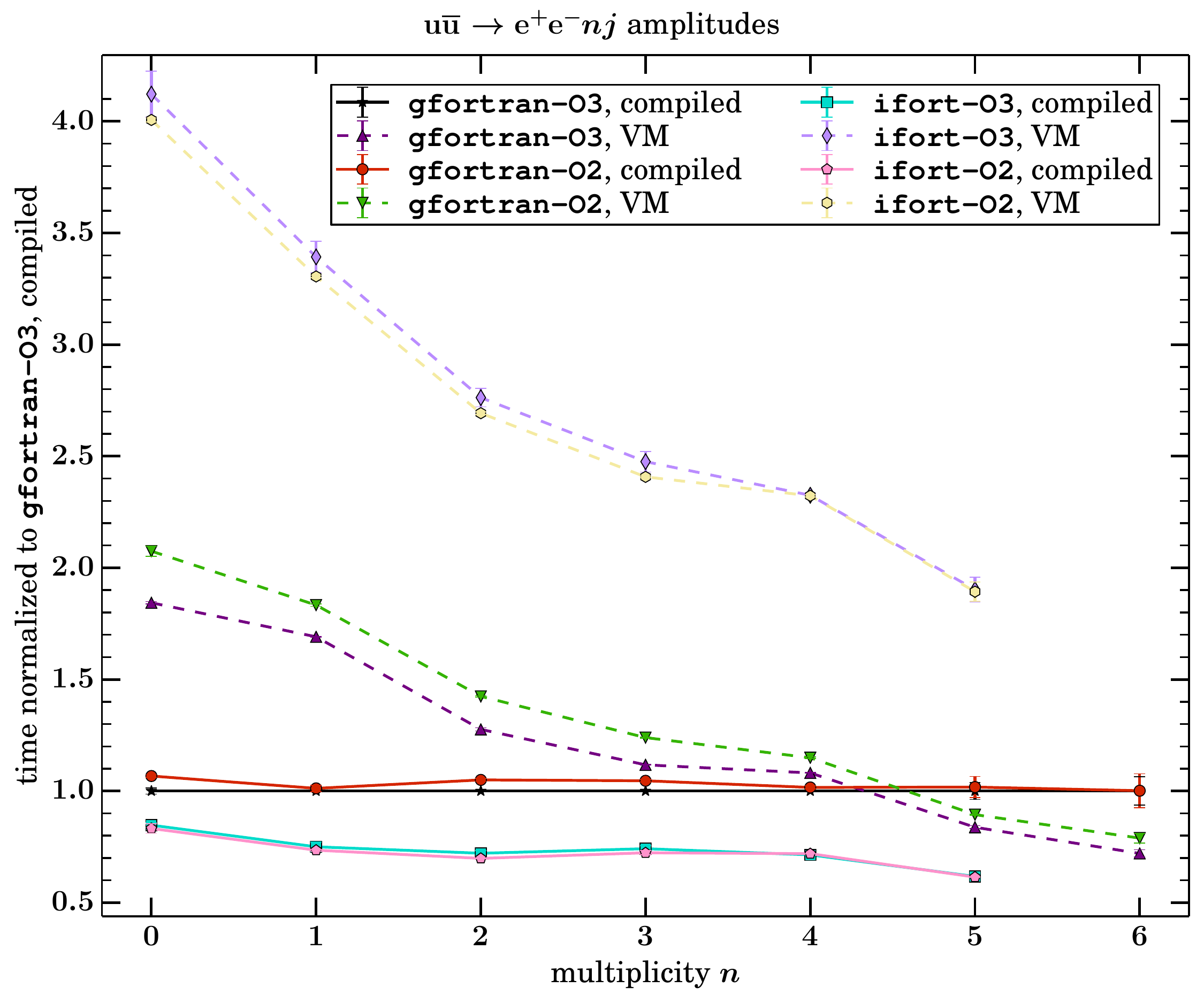}}
\caption{\ac{CPU} times measured with the \lang{Fortran} intrinsic
  \code{cpu\_time} and normalized for each process to the compiled source code
  using \code{gfortran -O3}.  Dashed (solid) lines represent the \ac{OVM}
  (compiled source code).}
\label{fig:runtime}
\end{figure}

To investigate the compiler dependence of the results, we use two different
compilers, \code{gfortran 4.7.1} and \code{ifort 14.0.3}.
We do not claim that our results are necessarily representative for all
\lang{Fortran} compilers or even compiler versions, but they should still give a
good impression of the expected variance in performance.
The evaluation time measurements are performed on a computer with two {Intel(R)
Xeon(R) E5-2440 @ 2.40GHz} \acp{CPU}, having \SI{16}{\mibi\byte} L3 cache on each
socket, and 2x \SI{32}{\gibi\byte} RAM running under {Scientific Linux 6.5}.
%
In Figure~\ref{fig:runtime}, we show the measured \ac{CPU} times for QCD and SM
processes with two different optimization levels for the compiled code and the
\ac{OVM} using the GNU and Intel compiler.
Since the evaluation times are highly reproducible, we use only three runs to
obtain mean and standard deviation.
The times are normalized for \emph{each} process to \cd{gfortran-O3}, which is
why the times are not growing with the number of particles.
For \cd{gfortran}, we observe for most processes the fastest performance
with \cd{-O3} and for \cd{ifort} with \cd{-O2}, which is an effect commonly
encountered.
The fastest performance is given by the source code compiled with \cd{ifort-O2}
being roughly \num{0.75} times the time needed by \cd{gfortran-O3}.
 
%
The crucial point, however, is that \cd{ifort} fails to compile the $n=7$ gluon
and the $\Pqu\Paqu\to\Ppositron\Pelectron6j$ Drell-Yan process while the \ac{OVM}
immediately starts computing.
The GNU compiler is usually able to compile one multiplicity higher compared
to the Intel before breaking down.
This fits together with the better performance of the compilable processes and
longer compile times as \cd{ifort} seems to apply more optimizations.

Another interesting observation is that the \ac{OVM} gets faster compared to the
compiled code with increasing multiplicity of external particles though this
feature is more pronounced in SM and QCD processes.
As discussed in \Rcite{1411.3834}, this is not due to costs from initialization 
or virtualization but likely caused by the explicit loop over the instructions
in the \ac{VM}.
It gives a higher probability to keep the decode function in the instruction
cache compared to the compiled code.
We observe roughly the same effect for both compilers, but the \ac{OVM}
compiled with \cd{ifort} is about a factor of two slower than the version
with \cd{gfortran}.
This is probably solvable with a profile-guided optimization, which allows the
compiler to use data instead of heuristics to decide what and how to optimize.
%
%
Amdahl's idealized law~\cite{amdahl} simply divides an algorithm into
parallelizable parts $p$ and strictly serial parts $1-p$. Therefore, the
possible speedup $s$ for a computation with $n$ processors is
\begin{equation}
  s(n) \equiv \frac{t(1)}{t(n)} = \frac 1{(1-p) + \frac p n}.
\label{amdahl}
\end{equation}
\begin{figure}[htb]
\subfigure[Pure gluon amplitudes $2\to(n-2)g$]{\includegraphics[width=0.49\textwidth]{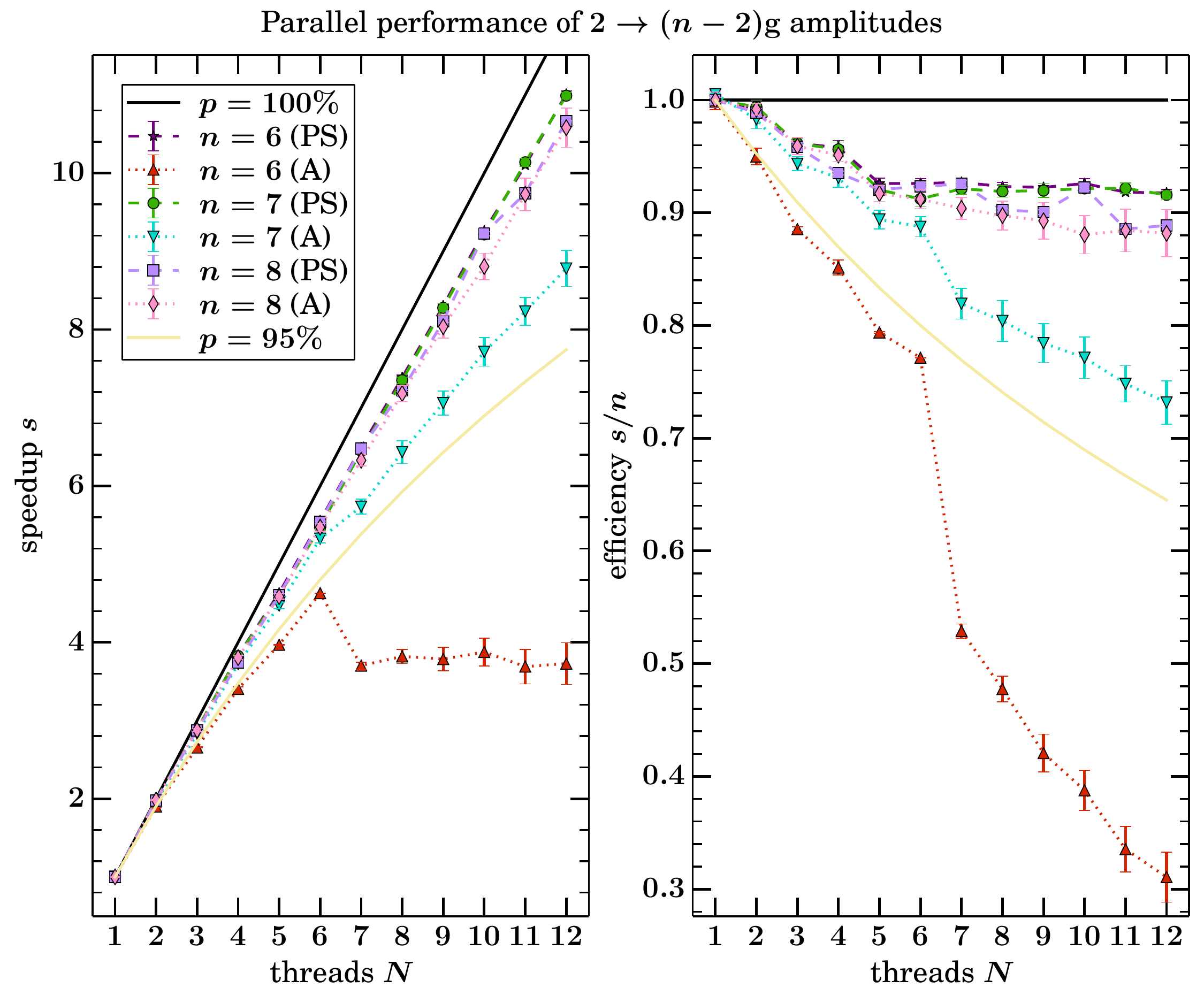}}\hfill
\subfigure[\ac{SM} Drell-Yan process $\Pqu \Paqu \to \Ppositron \Pelectron n j$]{\includegraphics[width=0.49\textwidth]{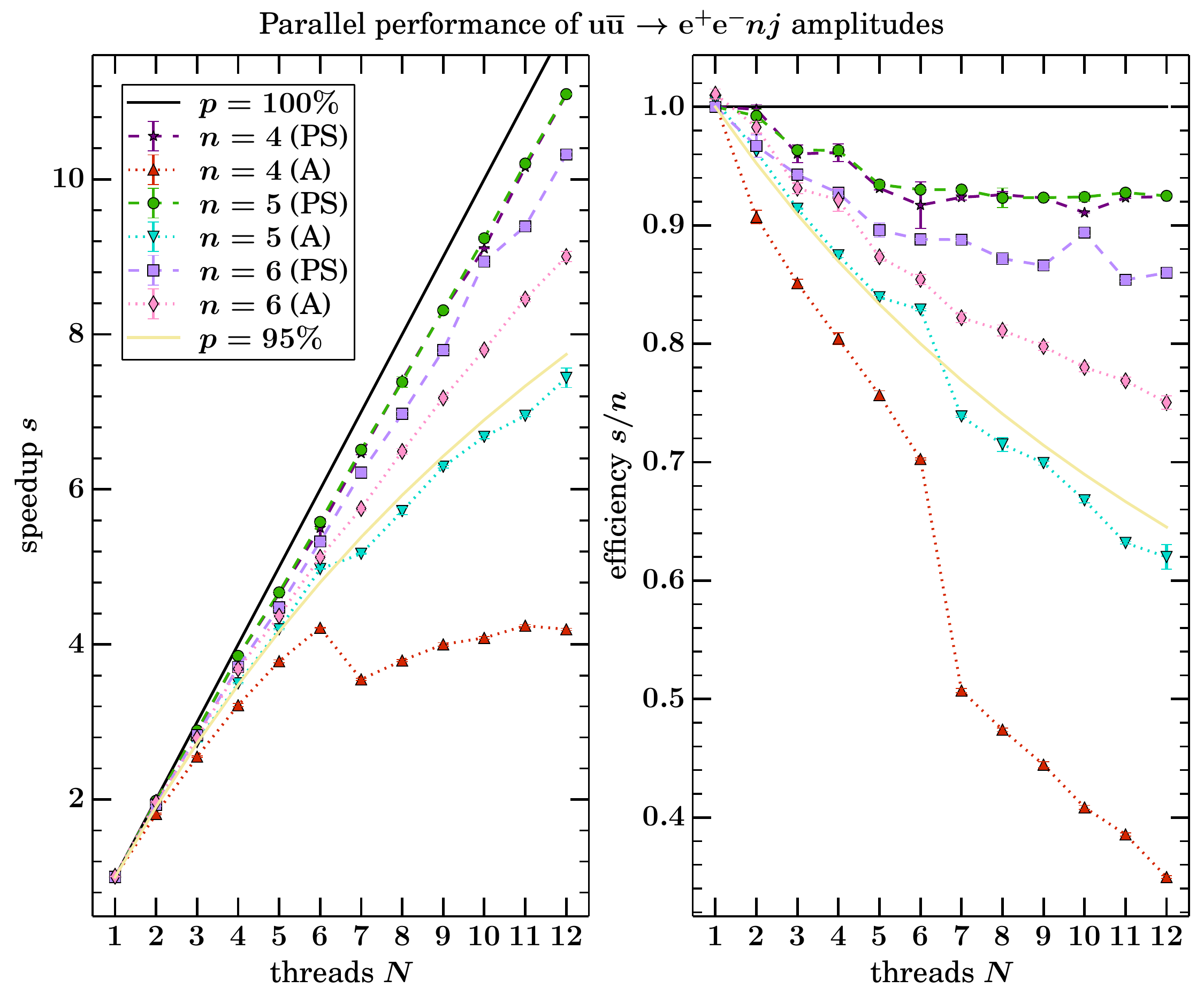}}
\caption{Speedup and efficiency to compute a fixed number of phase-space points
  for the parallel evaluation of multiple phase-space points (PS) and the
  parallel evaluation of the amplitude itself (A) are shown as dashed and dotted
  lines. The solid lines represent Amdahl's law for a fixed value of the
  parallelizable part $p$.}
\label{fig:parallel}
\end{figure}
Communication costs between processors $\order{n}$ have been neglected hereby in
the denominator of eq.~1.
This means that we have $\lim_{n\to\infty}s(n)=1/(1-p)$ in the idealized case and
$\lim_{n\to\infty}s(n)=0$ including communication costs.
In reality, we are interested in high speedups for finite $n$ and also have to
care about efficient cache usage.

In Figure~\ref{fig:parallel}, we
show the speedup with multiple cores, $N$, by either using the parallelization
procedure discussed above to compute one amplitude in parallel or by computing
multiple amplitudes for multiple phase-space points in parallel.
In Figure~\ref{fig:parallel} a), we can see that the $n=7$ and $n=8$ gluon amplitudes
parallelize very well with both methods with parallelizable parts above
$\SI{95}{\percent}$.
In the shared memory parallel evaluation of the amplitude (A), the impact of the
hardware architecture is quite obvious.
For $N=7$, \ie{} when the second socket of the board is activated, we see a drop
in efficiency.
This relative drop is the stronger the higher the communication costs are
compared to the calculation done in the individual threads and can thus be seen
most cleanly for the $n=6$ gluon and the $n=4$ Drell-Yan processes.
This is a cache coherency issue due to the synchronization of the caches of the
two CPUs.
%
%
We observe that the \bytecode{} for the \ac{OVM} is about one order of magnitude
smaller.
Some processes together with their compile times are shown in
Figure~\ref{tab:ovm_bc_creation}.
The smaller output format leads to less required RAM and time to produce it.
Especially for many color flows, where the generation time of \OMEGA{} is
dominated by the output procedure, we observe \eg{} for
$\Pgluon\Pgluon\to6\Pgluon$ a reduction in memory from \SI{2.17}{\gibi\byte} to
\SI{1.34}{\gibi\byte} and in generation time from
\SI{11}{\minute}~\SI{52}{\second} to \SI{3}{\minute}~\SI{35}{\second}, while
staying roughly the same for small processes.
\begin{table}[htbp]
  \caption{Size of the \bytecode{} (\cd{BC}) compared to the \cd{Fortran} source
    code together with the corresponding compile time with \cd{gfortran}. The
    compile times were measured on a computer with an \hardware{i7--2720QM}
    \ac{CPU}.
    The $2\Pgluon\to6\Pgluon$ process fails to compile.
    \vspace{1em} }
\label{tab:ovm_bc_creation}
  \centering
  \begin{tabular}{l c c c}
    \toprule{}
    process & \cd{BC} size & \cd{Fortran} size & $t_\T{compile}$ \\
    \midrule{}
    $\Pgluon \Pgluon \to \Pgluon \Pgluon \Pgluon \Pgluon \Pgluon \Pgluon$ &
    \SI{428}{\mibi\byte} & \SI{4.0}{\gibi\byte}
    & --- \\
    $\Pgluon \Pgluon \to \Pgluon \Pgluon \Pgluon \Pgluon \Pgluon$ &
    \SI{9.4}{\mibi\byte} & \SI{85}{\mibi\byte}
    & \SI{483+-18}{\second} \\
    $\Pgluon \Pgluon \to q \bar{q} q'\bar{q}' q''\bar{q}'' g$ & \SI{3.2}{\mibi\byte} &
    \SI{27}{\mibi\byte} & \SI{166+-15}{\second} \\
    $e^+ e^- \to5\,(e^+ e^-)$ &
    \SI{0.7}{\mibi\byte} & \SI{1.9}{\mibi\byte} & \SI{32.46+-0.13}{\second} \\
    \bottomrule{}
  \end{tabular}
\end{table}
\section{Summary and Outlook}
\label{s:conclusions}
%
A \ac{VM} circumvents the compile and link problems that are associated with
huge source code as it emerges from very complex algebraic expressions.
\acp{VM} are indeed a viable option that is maintaining relatively high
performance in the numerical evaluation of these expressions and allows to
approach the hardware limits.
In practice, a \ac{VM} saves at least hours of compile time.
The concept has been successively applied to construct the \ac{OVM} that is now
an alternative method to compute tree-level matrix elements in the publicly
available package \OMEGA{} and can be chosen in \whz{} with a simple
option since version 2.2.3.
Any computation can in principle be performed with a \ac{VM} though the benefits
are clearly in the regime of extreme computations that run into compiler limits
with the conventional method.
Here, we have seen that \acp{VM} can even perform better than compiled code.
Also the parallelization of the amplitude is for very complex processes close to
the optimum.

It would be an interesting experiment to remove the virtualization overhead by
using dedicated hardware that has the same instruction set as the \ac{OVM} to
compute matrix elements.
The number of instructions corresponding to different wave function
fusions and propagators is finite for renormalizable theories
(including effective theories up to a fixed mass dimension) and
implemented similarly in the various matrix element generators.  If the
authors can agree on a common set of instructions and conventions this
machine could therefore be used by all those programs.
\acp{FPGA} can serve as such a machine as they have comparable if not superior
floating-point performance with respect to current microprocessors and the
\ac{OVM} and its instruction set is the first step to test the feasibility and
potential gains of computing matrix elements in this environment.

\ack{}
JRR wants to thank the ACAT 2016 organizers for the beautiful venue in
Valpara\'{i}so and a fantastic conference in Chile. 
 
\printbibliography{}

\end{document}